\documentclass[final,3p,times,twocolumn]{elsarticle}
\usepackage{graphicx}
\usepackage{amsmath}
\usepackage{amssymb}

\journal{Computer Physics Communications}

\begin{document}

\begin{frontmatter}



\title{Quantum Annealing with Jarzynski Equality}


\author[label1]{Masayuki Ohzeki}
\author{Hidetoshi Nishimori$^b$}

\address[label1]{Department of Systems Science, Kyoto University, Yoshida-Honmachi, Sakyo-ku,
Kyoto 606-8501, Japan}
\address[label2]{Department of Physics, Tokyo Institute of Technology, Oh-okayama, Meguro-ku,
Tokyo 152-8551, Japan}
\begin{abstract}
We show a practical application of an well-known nonequilibrium relation, the Jarzynski equality, in quantum computation. 
Its implementation may open a way to solve combinatorial optimization problems, minimization of a real single-valued function, cost function, with many arguments. 
It has been disclosed that the ordinary quantum computational algorithm to solve a kind of hard optimization problems, has a bottleneck that its computational time is restricted to be extremely slow without relevant errors. 
However, by our novel strategy shown in the present study, we might overcome such a difficulty.
\end{abstract}


\end{frontmatter}



\section{Introduction: Quantum Annealing}

To reduce power loss in electric circuits, we have to minimize the circuit
length. This kind of problems are formulated into a more generic task to
minimize or maximize a real single-valued function of multivariables, cost
function. This is called optimization problem \cite{OP}. To solve these
problems is one of the most important tasks and has broad applications in
science and engineering. Well-known examples with discrete variables are
satisfiability problems, exact cover, maximum cut, Hamilton graph, and
traveling salesman problem.

Most of the above exemplified cases are mapped into a generic problem to
find the ground state for a kind of systems seen in statistical physics, spin glasses.
Its Hamiltonian is denoted as $H_0$ in the present study.
One of the generic algorithms to solve optimization problems in reasonable time by exploiting resources in physics is quantum
annealing (QA) \cite{QA1,QA2,QA3}. In QA, we introduce artificial degrees of
freedom of quantum nature, noncommutative operators in order to induce
quantum fluctuations. 
\begin{equation}
H(t) = f(t)H_0 + \left\{1- f(t)\right\} H_1,  \label{QAH}
\end{equation}
where $f(t)$ is assumed to be a monotonically increasing function satisfying 
$f(0) = 0$ and $f(\tau) = 1$. The annealing time is denoted by $\tau$. The
transeverse-field operator for spin glasses is often used as a quantum
fluctuation, $H_1 = - \sum_{i} \sigma_i^x$. The quantum annealing starts
from the ground state of $H_1$, a uniform linear combination of all spin
configurations on the basis of $\sigma_i^z$. The quantum system is driven by
gradually decreasing the quantum fluctuation according to $f(t)$. The
adiabatic theorem guarantees that we can reach a nontrivial ground state of $%
H_0$ after sufficiently slow quantum sweep $\tau_c \sim 1/\Delta^2$, where $%
\Delta$ is the energy gap of the instantaneous quantum system (\ref{QAH}) \cite{AD}. If
we consider the cases in which the quantum system as in Eq. (\ref{QAH}) has
a minimum energy gap vanishing as $\Delta \sim \exp(-\alpha N)$ for
increasing the system size $N$, QA does not work well in reasonable time 
\cite{FT,FT3}.

\section{Jarzynski Equality}

To overcome the above difficulty in hard optimization problems, we propose a
novel method in conjunction with a theoretical piece from non-equilibrium statistical physics, the Jarzynski equality (JE), in the
present study \cite{JE1,JE2}. The Jarzynski equality is written by an
well-known expression as, 
\begin{equation}
\left\langle \mathrm{e}^{-\beta W}\right\rangle =\frac{Z_{\tau }(\beta )}{%
Z_{0}(\beta )},
\end{equation}
where the angular brackets denote the average over all realizations in a
predetermined process starting from an initial equilibrium state and $W$ is
the work done during the process. 
The partition functions for the initial and final Hamiltonians are written as $Z_{0}(\beta )$ and $Z_{\tau }(\beta )$
with inverse temperature $\beta $, respectively. 
We here shortly recall the formulation of JE for the classical system on a heat bath. 
Let us consider a thermal nonequilibrium process in a finite-time schedule $t_0 = 0\leq t\leq t_n = n \delta t $. 
Thermal fluctuations can be simulated by the master equation. 
The probability that the system is in a state $\sigma _{k}$ at time $t_{k}$ is denoted as $%
P(\sigma _{k};t_{k})$. The transition probability per unit time $\delta t$
is defined as $M(\sigma_{k+1}|\sigma _{k};t_{k})$. In the original
formulation of JE, the work is defined as the energy difference merely
attributed to the change of the Hamiltonian, but we can construct JE also in
the case of changing the inverse temperature by defining the work as $-\beta
W(t_{k})=-(\beta (t_{k+1})-\beta (t_{k}))H_{0}$
for the state $\sigma $. The left-hand side of JE can then be expressed as 
\begin{eqnarray}
& & \left\langle \mathrm{e}^{-\beta W}\right\rangle  \notag \\
& & =\sum_{\{\sigma _{k}\}}\prod_{k=0}^{n-1}\left\{ \mathrm{e}^{-\beta
W(t_{k})}\mathrm{e}^{\delta tM(\sigma _{k+1}|\sigma
_{k};t_{k})}\right\}  \notag \\
&&\quad \times \tilde{P}(\sigma _{0};t_{0}),  \label{CJE}
\end{eqnarray}%
where $\tilde{P}(\sigma _{0};t_{0})$ denotes the initial equilibrium
distribution. Even if the transition term $\exp (\delta tM(\sigma
_{k+1}|\sigma _{k};t_{k}))$ is removed in this equation, JE is trivially
satisfied as one can simply confirm. A non-trivial feature of JE is in the
insertion of the transition term. From Eq. (\ref{CJE}), it is
straightforward to prove JE by use of the detailed-balance condition. An
observant reader may think the above formulation without any consideration
of quantum nature is not available for the application to QA. Nevertheless
we can apply the classical JE to QA by aid of the classical-quantum mapping 
\cite{QC}.

\section{Classical-quantum mapping}

The classical-quantum mapping leads us to a special quantum system, in which
the (instantaneous) equilibrium state of the above stochastic dynamics can
be expressed as the ground state. A general form of such a special quantum
Hamiltonian is given as $H_q(\sigma^{\prime }|\sigma;t) = I - \mathrm{e}^{%
\frac{\beta(t)}{2}H_0}M(\sigma^{\prime }| \sigma;t) 
\mathrm{e}^{-\frac{\beta(t)}{2}H_0}$. This Hamiltonian has the
ground state as $|\Psi_{\mathrm{eq}}(t)\rangle=\sum_{\sigma }\mathrm{e}%
^{-\beta (t)H_{0}/2}|\sigma\rangle/\sqrt{Z(t)}$. The ground state
energy is $0$, which can be explicitly shown by the detailed-balance
condition. On the other hand, the excited states have positive-definite
eigenvalues, which can be confirmed by the application of the
Perron-Frobenius theorem.

In the above special quantum system, we can deal with a quasi-equilibrium
stochastic process as an adiabatic quantum-mechanical dynamics in QA. Let us
consider QA for the above special quantum system by setting the parameter
corresponding to the temperature $T\to \infty$ ($\beta \to 0$). This
condition gives the trivial ground state for $H_1$ in the preceding section
with an uniform linear combination, similarly to the ordinary QA. If we
change $T\to 0$ very slowly, one can obtain the ground state of $H_q$,
which expresses the very low-temperature equilibrium state for $H_0$, the
cost function of the optimization problem that we wish solve. We however
consider to construct a protocol with the same spirit as JE by using the
special quantum system to overcome the bottleneck of the ordinary QA as
proposed in the following section.

\section{Quantum Jarzynski annealing and its application}

We prepare a trivial ground state with a uniform linear combination as the
initial condition in the ordinary QA. This initial state corresponds to the
high-temperature equilibrium state $|\Psi _{\mathrm{eq}}(t_{0})\rangle
\propto \exp (-\beta (t_{0})H_0/2)|\sigma \rangle $ with $\beta
(t_{0})\ll 1$. We introduce the exponentiated work operator $W_{\rm exp}(t_{k})=\exp (-(\beta (t_{k+1})-\beta (t_{k}))H_{0}/2)$. 
It looks like a non-unitary operator, but we can construct this operation by
considering an extended quantum system \cite{WQ,MO1}. 
If we apply $W_{\rm exp}(t_{k})$ to the preceding quantum state $|\Psi _{\mathrm{eq}%
}(t_{k})\rangle $, it is changed into a state corresponding to the
equilibrium distribution with the inverse temperature $\beta (t_{k+1})$.
After then, the time-evolution operator $U(\sigma ^{\prime }|\sigma
;t_{k+1})=\exp (-\mathrm{i}\delta tH_{q}(\sigma ^{\prime }|\sigma
;t_{k+1})/\hbar )$ also does not alter this state, since it is the ground
state of $H_{q}(\sigma ^{\prime }|\sigma ;t_{k+1})$. The resulting state
after the repetition of the above procedure is 
\begin{eqnarray}
&&|\Psi (t_{n})\rangle \propto \prod_{k=0}^{n-1}\left\{ W_{\rm exp}(t_{k})U_{k+1}(\sigma _{k+1}|\sigma _{k};t_{k})\right\} .  \notag \\
&&\quad \times |\Psi _{\mathrm{eq}}(t_{0})\rangle 
\end{eqnarray}%
This is essentially of the same form as Eq. (\ref{CJE}). We measure the
obtained state by the projection onto a specified state $\sigma ^{\prime }$.
The probability is then given by $|\langle \sigma ^{\prime }|\Psi
(t_{n})\rangle |^{2}$, which means that the ground state we wish to find is
obtained with the probability proportional to $\exp (-\beta (t_{n})H_{0})$.
If we continue the above procedure up to $\beta (t_{n})\gg 1$, we can
efficiently obtain the ground state of $H_{0}$. This is called the quantum
Jarzynski annealing (QJA) in the present study. Most of the readers, who are
familiar with the ordinary computation, have considered that it may seem
unnecessary to apply the time-evolution operator $U(\sigma _{k+1}|\sigma
_{k};t_{k})$, which expresses change between states by quantum fluctuations,
at the middle step between the operations of the exponentiated work
operators $W_{\rm exp}(t_{k})$. The time-evolution operator does not
mean an artificial control but describes the change by quantum nature during
quantum computation, which is inherent property in quantum computation.
However we here remember the nontrivial point of JE. Even if quantum nature
affects the instantaneous quantum state, the property of JE guarantees that
we keep the quantum state expressing the instantaneous equilibrium state. If
one considers to simulate this procedure in \textquotedblleft classical"
computers, we have to need the repetition of the pre-determined process to
deal with all fluctuations in the nonequilibrium-process average, since we
assume an ensemble in equilibrium in the formulation of JE. However,
when we implement QJA in \textquotedblleft quantum" computation, we operate
QJA to a single quantum system in principal since the classical ensemble is
mapped to the quantum wave function. We do not need the repetition of the
same procedure differently from the classical case \cite{MO1}.

Let us take a simple instance to search the minimum from a one-dimensional
random potential, which is formulated as the Hamiltonian $%
H_{0}=-\sum_{i=1}^{N}V_{i}|i\rangle \langle i|$. Here $V_{i}$ denotes the
potential energy at site $i$ and chosen randomly. We employ a linear
schedule for tuning the parameter $\beta $ from $0$ to $100$. Figure \ref%
{Plot} shows the plot for QJA (upper circles), which are fixed along the
reference curves (solid curve) representing the instantaneous
Gibbs-Boltzmann factor. In contrast, QA (lower triangles) can not
sufficiently find the ground state since we consider a very short annealing
is considered in this case. 
\begin{figure}[tbp]
\begin{center}
\includegraphics[width=70mm]{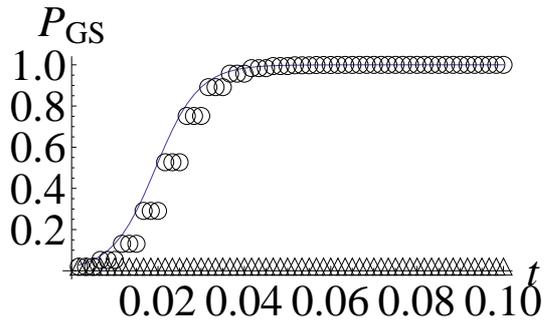}
\end{center}
\caption{{\protect\small The performance of QA (triangles), and QJA
(circles). The reference curve describes the instantaneous Gibbs-Boltzmann
factor. }}
\label{Plot}
\end{figure}

\section{Summary}

We consider an application of JE to quantum computation as QA to solve the
optimization problems by using the classical-quantum mapping. As we
expected, this protocol keeps the quantum system to express the equilibrium
state for the instantaneous inverse temperature. The result by QJA shown
here gives the ground state in a short annealing and implies that we may
overcome the difficulties in hard optimization problems and solve them in a
reasonable time. The present result is nothing but preliminary one. We should
address the problem on practical efficiency for several interesting hard
problems we wish to solve in the future study \cite{MO2}.

\section*{Acknowledgement}

This work was supported by CREST, JST.





\end{document}